\documentclass[%
 reprint,
 amsmath,amssymb,
 aps,
]{revtex4-2}

\usepackage{graphicx}
\usepackage{dcolumn}
\usepackage{bm}
\usepackage{xcolor}

\begin{document}

\title{Improved ionization potential of calcium using frequency-comb based Rydberg spectroscopy}

\author{Chanhyun Pak}
\author{Matthew J. Schlitters}
\author{Scott D. Bergeson}%
\email{scott.bergeson@byu.edu}
\affiliation{Department of Physics and Astronomy, Brigham Young University, Provo, Utah 84602, USA}

\date{\today}

\begin{abstract}
We report new frequency-comb-based measurements of Ca Rydberg energy levels. Counter-propagating laser beams at 390 nm and 423 nm excite Ca atoms from the $4s^2~^1\mbox{S}_0$ ground state to $4sns~^1\mbox{S}_0$ Rydberg levels with $n$ ranging from 40 to 110. Near-resonant two-photon two-color excitation of atoms in a thermal beam makes it possible to eliminate the first-order Doppler shift. The resulting lineshapes are symmetric and Gaussian. We verify laser metrology and absolute accuracy by reproducing measurements of well-known transitions in Cs, close to the fundamental wavelengths of our frequency-doubled ti:sapphire lasers. From the measured transition energies we derive the ionization potential of Ca, $E_{\rm IP} = 1, 478, 154, 283.42 \pm 0.08 \mbox{(statistical)} \pm 0.07 \mbox{(systematic)}$ MHz, improving the previous best determination by a factor of 11.
\end{abstract}

\maketitle

\section{Introduction}

Spectroscopy of Rydberg atoms finds applications in a number of areas. These include quantum information \cite{Saffman2016, Madjarov2020, doi:10.1116/5.0036562}, detection of rare isotopes \cite{Mostamand2022}, atomic structure calculations \cite{vaillant2012, Berl2020}, RF field detection \cite{Jing2020, Holloway2021}, and many-body dynamical studies \cite{Camargo2018, Shaffer2018, Qiao2021}. These applications, as well as many others, exploit the Rydberg atoms' extreme sensitivity to electric and magnetic fields.

In a low-density, field-free environment, the energy level structure of unperturbed Rydberg atoms is  straightforward. The Ritz formula \cite{Langer1930} predicts energy levels,
\begin{equation}
  E_n = E_{\rm IP} - \frac{R}{[n-\delta(n)]^2},
  \label{eqn:rydberg}
\end{equation}
where $E_n$ is the energy of the level with a principle quantum number of $n$, $E_{\rm IP}$ is the ionization potential of the atom, $R$ is the Rydberg constant with a finite mass correction for the atom, and $\delta(n)$ is the quantum defect. The pure power series expansion of $\delta(n)$ is
\begin{equation}\label{eqn:quantum_defect}
  \delta_n = \delta_0 + \frac{\delta_1}{(n-\delta_0)^2} +
  \frac{\delta_2}{(n-\delta_0)^4} + \cdots,
\end{equation}
where the quantum defect parameters $\delta_0, \delta_1, \delta_2$ can be calculated using multi-channel quantum defect theory or fit using experimental data \cite{Gentile1990}.

In calcium atoms, the quantum defects for $S$, $P$, and $D$ singlet and triplet states are known with high accuracy in the range of $n=22-55$ \cite{Gentile1990}. This work was extended to the range of $n=20-150$ for singlet $P$ and $F$ states \cite{Miyabe2006}. The quantum defects for singlet $S$ states has also been confirmed within the range of $n=40-120$ \cite{Zelener2019}.

A recent publication reported calcium Rydberg energy level measurements in a magneto-optical trap (MOT) \cite{Zelener2019}. A near-resonant laser at 423 nm together with a near-UV laser excited the calcium MOT atoms to Rydberg states. The laser frequencies were determined using a calibrated wavelength meter, similar to other studies \cite{Saakyan2015,Saleh2015,Cout2018,Konig2020,Orson2021}. On resonance, the two-photon excitation opened a loss channel in the MOT and the steady-state MOT fluorescence decreased. Measuring the MOT fluorescence as a function of laser frequency produced an asymmetric lineshape, due to Rydberg-Rydberg atom interactions in the MOT \cite{vaillant2012}. In that study, no lineshape analysis was given, even though atom-atom interactions can shift the line center \cite{Pillet2016}. Systematic uncertainties in determining the line center were 10 MHz. The study used Eq.~(\ref{eqn:rydberg}) to produce a value of $E_{\rm IP}$ with an uncertainty of 1.2 MHz, consistent with previous less accurate work \cite{Miyabe2006}. The accuracy of such a conclusion is based on assumptions about the nature of the experimental and statistical uncertainties, and neglects systematic errors such as line-shape, AC Stark shift, and density shifts.

In this paper we present measurements of calcium Rydberg energy levels and a determination of the Ca ionization potential. We describe our Doppler-free atomic spectroscopy and laser metrology together with a detailed error analysis. We report an ionization potential of $E_\text{IP} = $ 1,478,154,283.42 $\pm$ 0.11 MHz (statistical and systematic uncertainties combined), improving previous measurements by a factor of 11.

\section{Experiment}
\label{sec:experiment}

\begin{figure}
\includegraphics[width = \columnwidth]{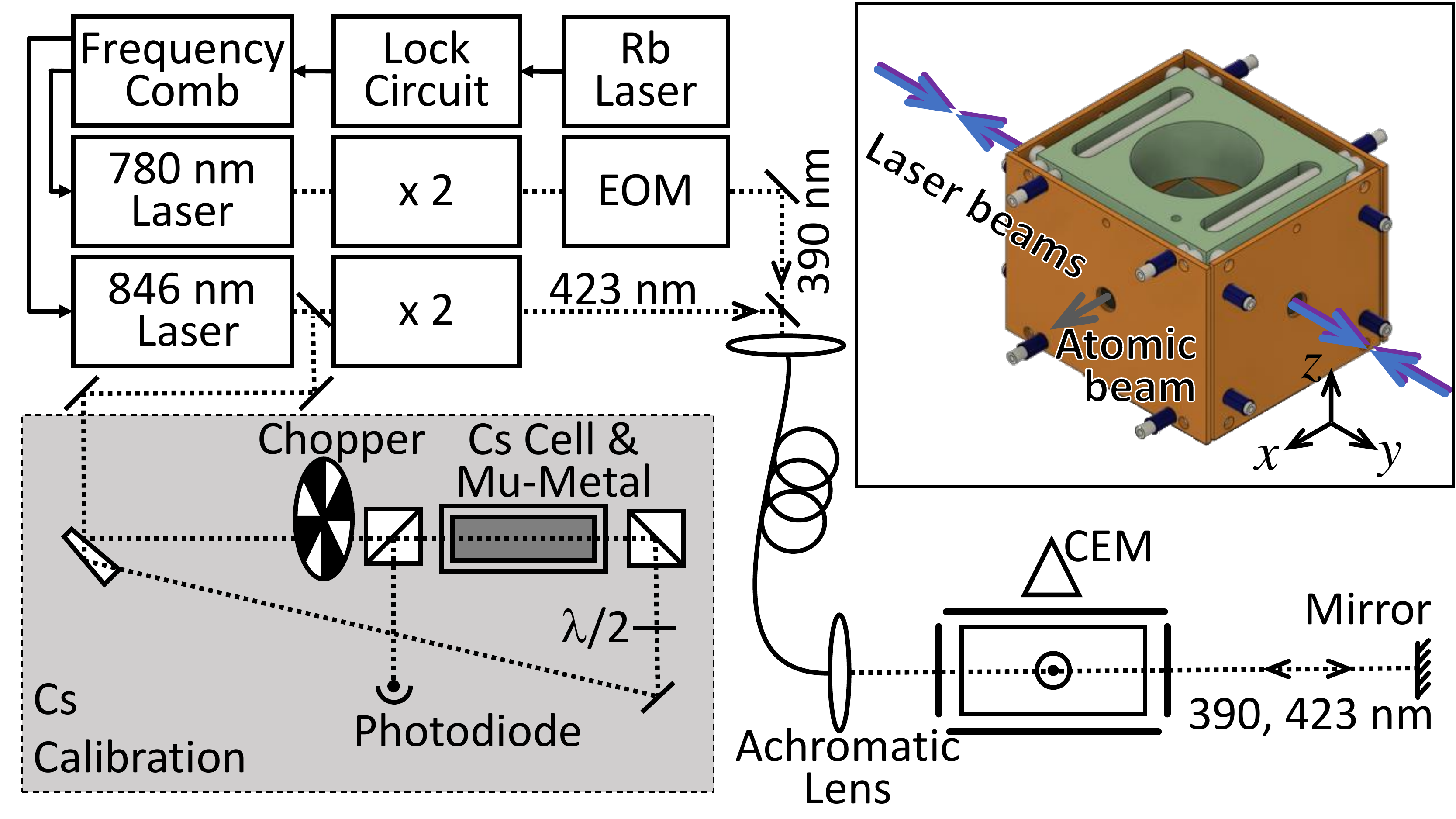}
\caption{\label{fig:experimental_setup} Schematic diagram of experiment setup. Continuous-wave  ti:sapphire lasers at 780 and 846 nm are offset-locked to a GPS-disciplined ti:sapphire frequency comb. The lasers are frequency doubled to 390 and 423 nm and coupled to a single-mode polarization-maintaining optical fiber. The collimated laser beams cross a collimated Ca atomic beam to excite atoms to $4sns\;^1\mbox{S}_0$ levels. The shaded region at the lower left shows the Cs saturated absorption setup used to verify our laser metrology. CEM = channel electron multiplier. Inset: A CAD rendering of the interaction region. The box is made entirely from copper, held together using alumina spacers and metal springs. Not shown is mesh across the bottom of the top plate, allowing ions to pass through to the time-of-flight tube and ion detector. The 390 and 423 nm laser beams both co-propagate and counter-propagate across the atomic beam.}
\end{figure}

\begin{figure}
\includegraphics[width = 0.95\columnwidth]{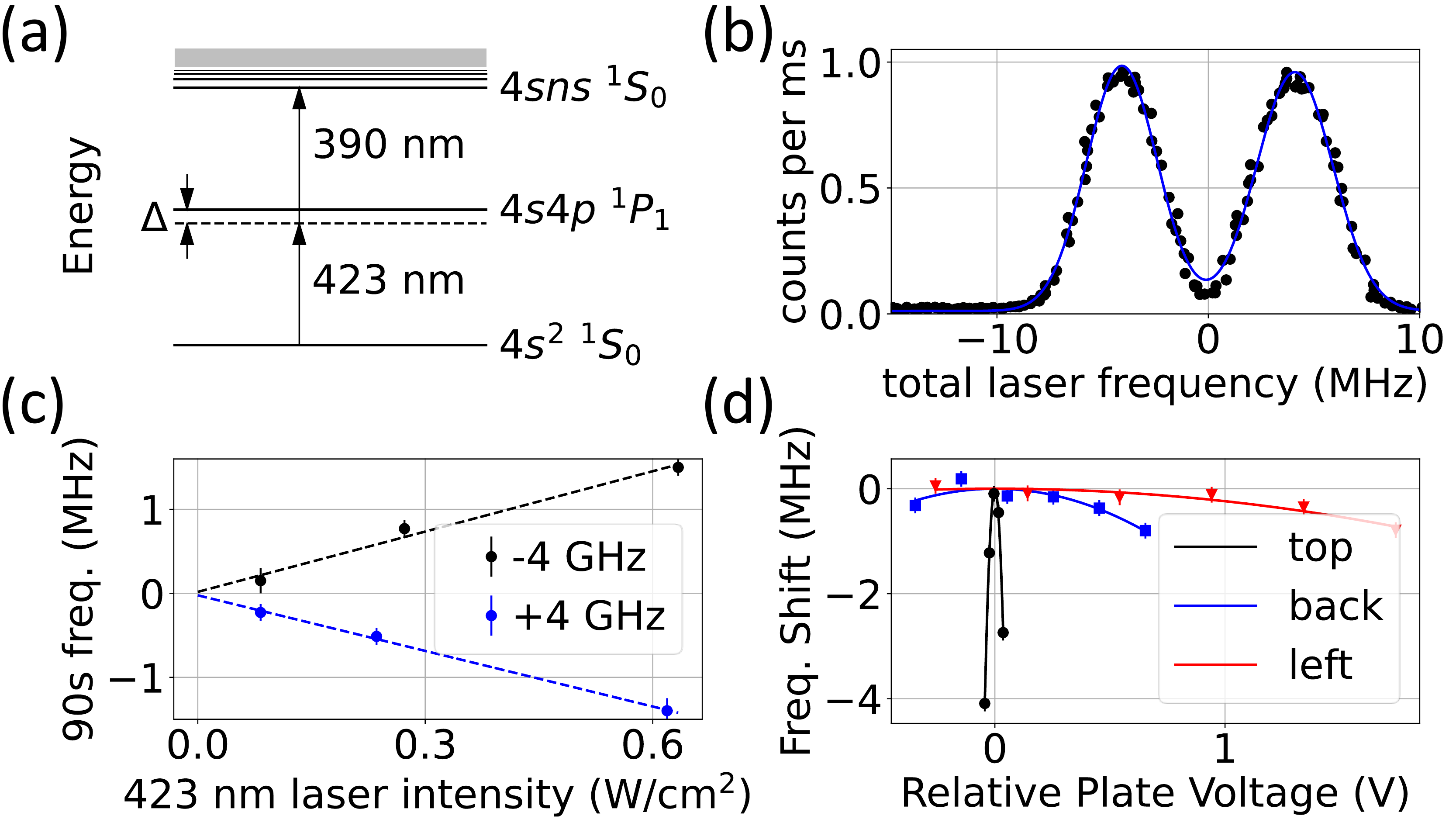}
\caption{\label{fig:figures} Experimental details. (a) Partial energy level diagram for Ca showing the intermediate state detuning, $\Delta$. (b) Typical 90s Rydberg spectrum. Two Doppler-shifted peaks appear while scanning the 423 nm laser as discussed in the text. (c) The AC Stark shifts vs. 423 nm laser power, extrapolated to zero power. The extrapolations with  $\Delta/2\pi = \pm$4000 MHz agree with each other. (d) DC Stark shift minimization.}
\end{figure}

A schematic diagram of our experimental setup is shown in Fig.~\ref{fig:experimental_setup}. An atomic beam is created by heating calcium to 500 $^{\circ}$C. The beam is collimated using an aligned microcapillary array, similar to previous work \cite{senaratne2015effusive,sprenkle2022}. Calcium atoms are excited to Rydberg states using counter-propagating lasers, one with a wavelength of 423 nm and a second laser at wavelengths between 390 and 391 nm. The 423 nm laser is detuned from the Ca $4s^2~^1\mbox{S}_0 - 4s4p~^1\mbox{P}_1$ transition by typically -4000 MHz as shown in Fig.~\ref{fig:figures}(a).

We use frequency-doubled ti:sapphire lasers to generate laser radiation at 423 nm and near 390 nm. The two laser beams are overlapped on a dichroic mirror and coupled into a large-area single-mode optical fiber. The output of the fiber is collimated using an achromatic lens. After crossing the atomic beam, the lasers are retroreflected back into the optical fiber to ensure nearly perfect counter-propagating alignment, as shown in Fig.~\ref{fig:experimental_setup}. The 390 nm laser beam is pulsed using an electro-optic modulator (EOM). The optical pulse width is approximately 6.5 $\mu$s, FWHM.

Three pairs of copper plates connected by shielded alumina rods and spacers are used to eliminate the DC Stark shift (see inset to Fig. \ref{fig:experimental_setup}), similar to many other studies, for example \cite{Ling1992}. We ionize the Rydberg atoms by applying a capacitively-coupled, $1~\mu$s duration, -100 V pulse onto the top copper plate. The pulse rises 18.5 $\mu$s after the peak of the 390 nm laser pulse. The pulse directs ions towards the time-of-flight tube and channel electron multiplier detector. The excitation-ionization cycle repeats at a rate of 10 kHz. We scan the 423 nm laser over a frequency range of $\pm 13$ MHz. At each laser frequency setting, we count the number of detected ions during a 4 second time interval [see Fig.~\ref{fig:figures}(b)].


The excitation lasers are not exactly the same wavelength. The first-order Doppler shift is not automatically eliminated. Both laser beams counter-propagate across the atomic beam in both directions. Because the laser propagation direction is not exactly perpendicular to the atomic beam, two Doppler-reduced peaks are observed. They are symmetrically shifted above and below the Doppler-free frequency. The first-order Doppler shift depends on the laser $k$-vector, $|\vec{k}|=2\pi/\lambda$, and the thermal velocity of the calcium atoms, $v=({k_{\rm B} T/ m})^{1/2} = 400~\mbox{m/s}$,
\begin{equation}
    \Delta f  = \frac{1}{2\pi}\sum \vec{k} \cdot \vec{v},
\end{equation}
The angle between $\vec{k}$ and $\vec{v}$ is $\theta = 87^{\circ}$. The resulting Doppler shifts are,
\begin{align}
    \Delta f & = \pm \left(
      400~\mbox{m/s}
    \right)
    \cos(\theta)
    \left(
      \frac{1}{390~\mbox{nm}} - \frac{1}{423~\mbox{nm}}
    \right) \\
    & = \pm 3.5 \text{ MHz}, \nonumber
\end{align}
where the $\pm$ comes from the relative directions of the 390 nm and 423 nm laser beams. We fit the ion count rate vs. laser frequency to a two-Gaussian lineshape function. Averaging the center frequency of the two Gaussian functions gives the Doppler-free transition frequency. Typical data for the $90s$ state are shown in  Fig.~\ref{fig:figures}(b).

The apparent transition frequency depends on the excitation laser power, as shown in Fig.~\ref{fig:figures}(c). Our laser intensities range up to 600 mW/cm$^2$ 1100 mW/cm$^2$ for the 423 and 390 nm lasers respectively. The Rabi frequency for the $4s^2~^1\mbox{S}_0 - 4s4p~^1\mbox{P}_1$ transition is $\Omega_{423} = \gamma (I/2I_{\rm sat})^{1/2} = 2\pi \times 78~\mbox{MHz}$, where $\gamma=2\pi \times 35~\mbox{MHz}$ for the 423 nm transition and $I_{\rm sat} = 60~\mbox{mW/cm}^2$. Based on scaling the results of Ref. \cite{DEIGLMAYR2006293}, we estimate the Rabi frequency for the $4s4p~^1\mbox{P}_1 - 4s90s~^1S_0$ transition to be $\Omega_{390} = 2\pi \times 12$ kHz. The two-photon Rabi frequency is $\Omega_{423} \Omega_{390} / \Delta = 2\pi \times 0.2$ kHz.

We use a frequency comb to determine the frequencies of our lasers, as described in previous publications \cite{Lyon:14, kleinert2016measurement, mcknight2018}. One mode of the frequency comb is offset-locked to a Rb-stabilized diode laser. The GHz comb repetition rate is not stabilized but varies typically 1 Hz in a 1-second measurement interval. Other lasers in the experiment are offset-locked to their nearest comb modes, and the beat notes between the lasers and their nearest comb modes are measured. The laser frequency is calculated using,
\begin{equation}
    f_{\rm laser} = f_{\rm Rb} \pm f_{\rm Rbb} + Nf_{\rm rep} \pm f_{\rm beat},
    \label{eqn:comb}
\end{equation}
where the $f_{\rm Rb}$ is the $^{87}$Rb $D_2~F = 2 \rightarrow F' = (2,3)$ crossover transition frequency of the Rb laser \cite{Ye:96}, $f_{\rm Rbb}$ is the beat note between the Rb-stabilized diode laser and the nearest frequency comb mode, $f_{\rm beat}$ is the beat note between the 846 nm or 780 nm laser and the nearest frequency comb mode, and $N$ is an integer.

Excitation of the Rydberg level uses the second harmonics of the 846 nm and 780 nm lasers, both of which are locked to the frequency comb. Using Eq.~\eqref{eqn:comb}, the Rydberg transition frequency is calculated to be,
\begin{equation}
    f_{\rm Ryd}  = 2\left[
    2\left(
    f_{\rm Rb} \pm f_{\rm Rbb}
    \right)
    - N_2 f_{\rm rep} \pm f_{\rm 846b} \pm f_{\rm 780b}
    \right],
    \label{eqn:ryd-laser}
\end{equation}
where $f_{\rm 846b}$ and $f_{\rm 780b}$ are the beat notes between the two infrared lasers and the frequency comb, and $N_2$ is an integer. The ambiguity in the beat note signs is removed by measuring the change in the un-locked beat note when we change the laser frequency.

\section{Error analysis}

\begin{table}
\caption{
\label{tab:error}
Statistical uncertainties in laser metrology and atomic spectroscopy. All values are given in terms of their contribution to the total Rydberg frequency measurement.
}
\begin{ruledtabular}
\begin{tabular}{lr}
\multicolumn{2}{c}{Error contribution (MHz)}\\
\colrule
DC Stark shift &  0.130 \\
AC Stark shift & 0.100\\
Comb $f_0$ & 0.200\\
$f_{\text{rep}}$ & 0.012\\
423 nm (846 nm) laser lock to comb & 0.060 \\
390 nm (780 nm) laser lock to comb & 0.060 \\
Rb laser lock to comb & 0.120 \\\hline
Total & 0.300 \\
\end{tabular}
\end{ruledtabular}
\end{table}

The errors can be divided into two classes: atomic spectroscopy and laser metrology. Table~\ref{tab:error} lists a summary of our error analysis. We minimize the DC stark shift by applying small voltages to the copper plates surrounding the excitation region. The apparent transition frequency shifts quadratically in applied voltage, as shown in Fig.~\ref{fig:figures}(d). We fit the frequency-vs.-voltage measurements to a parabola and acquire data at voltages minimizing the Stark shift. The statistical uncertainty in the transition frequency due to the DC Stark shift is typically 0.13 MHz.

To eliminate the AC Stark shift, we measure the Rydberg transition frequency for a range of 423 nm laser power, and for two values of the intermediate state detuning. We extrapolate the measured transition frequencies to zero laser power, as shown in Fig.~\ref{fig:figures}(c). The uncertainty in the extrapolated value is typically 0.1 MHz at the Rydberg transition frequency.

Because the ground and Rydberg states have no angular momentum, the Zeeman shift vanishes. We address the Doppler shift in Sec. \ref{sec:experiment}. In our measurements, the angle between the atomic beam and the laser beams is $\theta=87^{\circ}$. The line shapes are symmetric with widths that indicate the frequency noise in our excitation lasers.

The frequency of the Rydberg transition, Eq.~\eqref{eqn:ryd-laser}, depends on the $^{87}$Rb $D_2~F = 2 \rightarrow F' = (2,3)$ crossover transition, the lock point of our reference laser relative to that frequency, and several beat note frequencies in the 20 to 1000 MHz range.

While the Rb transition is well-known, our frequency lock relative to that transition could be shifted due to offsets in the locking electronics. To verify the accuracy of the frequency comb, including the reference laser offset relative to the Rb transition, we measure the well-known Cs $D_2~F = 4 \rightarrow F' = (4,5)$ crossover transition  \cite{Gerginov2004}. Typical data is shown in Fig.~\ref{fig:figures}(d). Uncertainties in determining this offset are estimated by measuring all 6 lines in the Cs $D_2$ hyperfine array relative to their known values, and also by repeated measurements of the Cs $D_2~F = 4 \rightarrow F' = (4,5)$ crossover transition. We conservatively estimate the statistical uncertainty in the lock point to be 0.05 MHz at 852 nm. It is quadrupled [see Eq.~\eqref{eqn:ryd-laser}] and included in Table~\ref{tab:error} as ``Comb $f_0$''.

The beat note frequencies are measured using RF spectrum analyzers. The analyzers are referenced to a 10 MHz Rb oscillator. The Rb oscillator is referenced to the GPS satellite system, phase-locked using the 1 PPS output from a GPS receiver on the roof of our building. The frequency error in this system is negligible for our measurements.

In our measurements, the 846 nm laser is scanned. The 846 nm laser is tightly offset-locked to the frequency comb. The $f_{\rm 846b}$ beat note spectrum is averaged for 4 seconds per measurement point. The peak of the averaged beat note spectrum is used for $f_{\rm 846b}$. The repeatability in determining the beat note is approximately 30 kHz. The  $f_{\rm 780b}$  and  $f_{\rm Rbb}$ beat note spectra are also averaged for 4 seconds and read out the same way. However, because these laser frequencies do not change during the measurement process, the 4-second measurements are averaged together during one scan over the Rydberg transition. The statistical uncertainty in determining these beat notes is also approximately 30 kHz. This uncertainty is doubled for the 846 nm and 780 nm lasers and quadrupled for the Rb laser in Table \ref{tab:error}.

The frequency comb repetition rate,  $f_{\rm rep} = 0.988~\mbox{GHz}$, is measured using a GPS-referenced frequency counter. The counting interval is 0.5 seconds, read out 8 times during the 4-second measurement time. The standard deviation of these measurements is typically 0.5 Hz. Assuming a normal distribution, the uncertainty in the mean is typically 0.2 Hz. The integer $N_2$ in Eq.~\ref{eqn:ryd-laser} is roughly 30,000, making the uncertainty in  $f_{\rm rep}$ equal to 6 kHz in a 4-second measurement time. This statistical uncertainty is doubled in Table \ref{tab:error}.

\begin{table}
\caption{\label{tab:energy} Table of $4s^2\; ^1\mbox{S}_0 - 1sns\; ^1\mbox{S}_0$ Rydberg transition energies and calculated ionization potential energies. Values of $E_{\rm IP}$ are calculated using Eqs.~\eqref{eqn:rydberg} and \eqref{eqn:quantum_defect} and quantum defect results from Ref. \cite{Gentile1990}.}
\begin{ruledtabular}
\begin{tabular}{c c c}
\multicolumn{1}{c}{$n$}&
\multicolumn{1}{c}{ measured $E_n$ (MHz)}&
\multicolumn{1}{c}{$E_I$ (MHz)}\\
\colrule
40  & 1475834972.82 & 1478154283.09\\
50  & 1476706104.14 & 1478154283.52\\
60  & 1477164846.49 & 1478154283.39\\
70  & 1477435698.36 & 1478154283.44\\
80  & 1477608838.11 & 1478154283.08\\
90  & 1477726183.13 & 1478154283.38\\
100 & 1477809364.80 & 1478154283.80\\
110 & 1477870463.33 & 1478154283.69\\
\hline
Average &  & 1478154283.42\\
Stand. dev. &  & 0.24 \\
Unc. in mean & & 0.08 \\
Systematic unc. & & 0.07
\end{tabular}
\end{ruledtabular}
\end{table}

\begin{figure}[t!]
\includegraphics[width=\columnwidth]{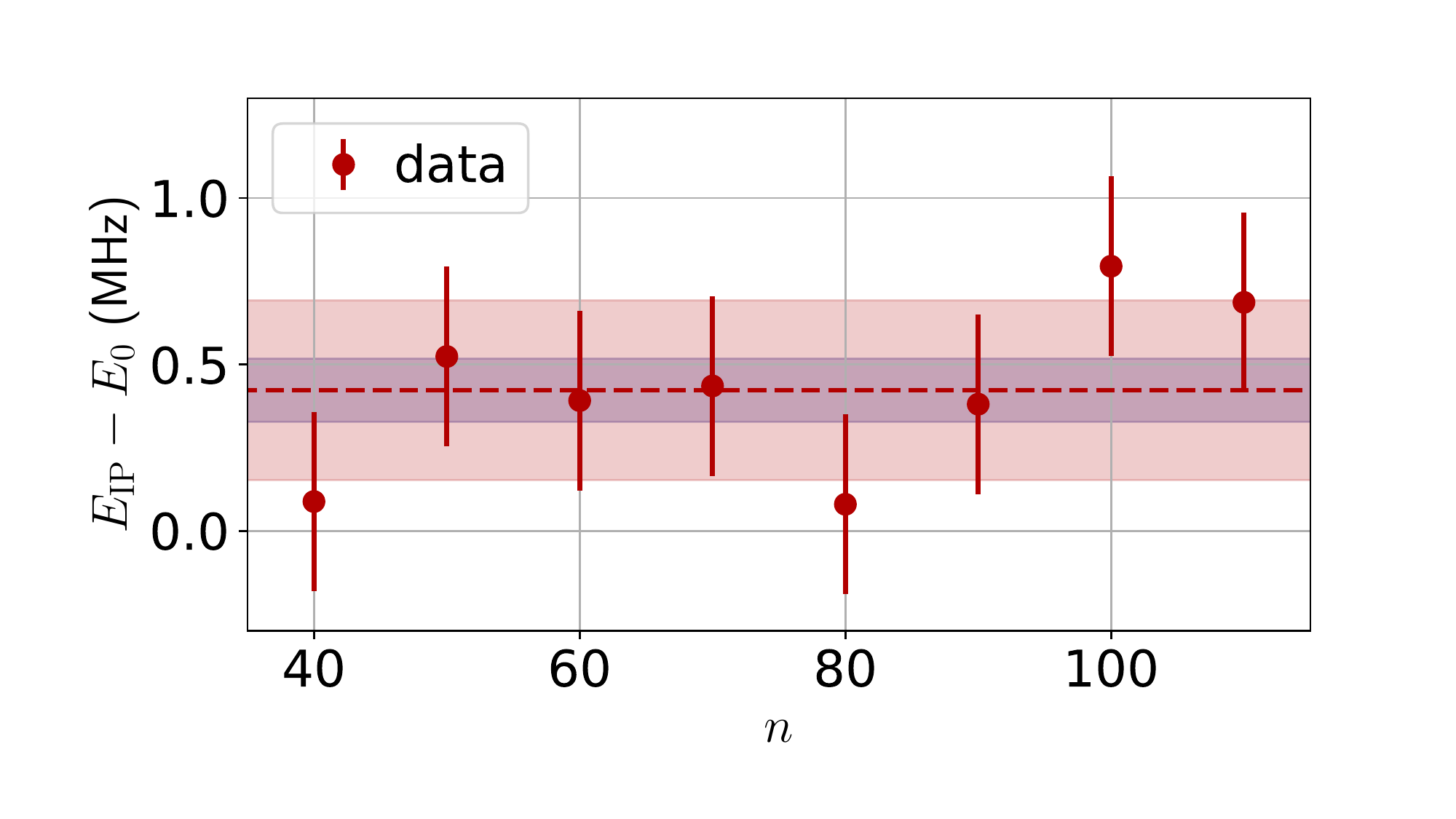}
\caption{\label{fig:Eip_plot} A plot of $E_{\rm IP}$ vs. principal quantum number $n$. The error bars for the present work are from Table~\ref{tab:error}. The shaded red area indicates the standard deviation in the calculated $E_{\rm IP}$ values. The gray shaded area indicates the uncertainty in the mean.}
\end{figure}

All cell-based frequency measurements must allow for shifts due to possible impurities in the cell. In the work of Refs.~\cite{Wu2013, Wu2015,chen2020influence}, it was shown that the Cs $6s(F=4) - 8s(F^{\prime}=4)$ transition shifts linearly with He pressure. They showed that the dominant impurity in aging Cs cells is atmospheric He diffusing through the glass wall. They also showed that the transition linewidth can be used as a surrogate for cell impurity measurements when the laser linewidth is well-characterized. We have measured this transition in our Cs reference cell. Our zero-power extrapolated FWHM is 1.06 MHz. For this measurement, special efforts were made to minimize the noise in the frequency comb. In our $6s-8s$ measurements, transit time broadening accounts for approximately 40 kHz of the observed linewidth. Using the data from Fig.~3 of Ref.~\cite{chen2020influence}, and assuming zero contribution from the laser linewidth, a transition FWHM of 1.02 MHz corresponds to a shift of 18 kHz. While the pressure shift of the $D_2~F = 4 \rightarrow F' = (4,5)$ crossover transition is not known, we can use this 18 kHz as an estimate of the systematic uncertainty in the $f_{\rm Rb}$ determination in Eq.~\eqref{eqn:ryd-laser}.

\section{Results and Discussion}

Our measured frequencies for the Ca $4s^2~^1\mbox{S}_0 - 4sns~^1\mbox{S}_0$ transitions with $n = 40-110$ are listed in Table~\ref{tab:energy} and plotted in Fig.~\ref{fig:Eip_plot}. For each Rydberg level measurement, we calculate the ionization energy using the Ritz formula of Eq.~\eqref{eqn:rydberg}, with the quantum defect expression in  Eq.~\eqref{eqn:quantum_defect}. The quantum defect values $\delta_0 = 2.33793016(300)$, and $\delta_1 = -0.114(3)$, and the Rydberg constant $R_{\rm Ca} = 109735.809284$ cm$^{-1}$ are from Ref. \cite{Gentile1990}. The standard deviation in the ionization energy calculations is 0.24 MHz, in close agreement with the estimated statistical uncertainties in Table~\ref{tab:error}. Assuming that these Ionization Energy uncertainties are statistical and uncorrelated, the estimated error in the mean is $\sigma_\mu = \sigma / N^{1/2} = 0.24 /8^{1/2} = 0.08$ MHz. We add the estimated systematic uncertainty of $4 \times 0.018~\mbox{MHz} = 0.072~\mbox{MHz}$, the estimated uncertainty due to pressure shifts in the Cs cell, in quadrature for a final estimated uncertainty of 0.11 MHz in $E_{\rm IP}$.

Our results can be compared to a recent publication \cite{Zelener2019}. Their work differs from ours in that they used the steady-state number of atoms in a magneto-optical trap as their ``detector'' and measured their laser frequency using a laser wavelength meter. They also use a modest intermediate state detuning of -47.8 MHz compared to our value of -4000 MHz. Their typical transition linewidth was asymmetric with a full width of approximately 10 MHz.
Coherent Rydberg excitation at low values of the intermediate state detuning are broadened by the intermediate-state decoherence. Excitation line shapes can become asymmetric due to the Autler-Townes splitting \cite{DEIGLMAYR2006293} and atom-atom interactions \cite{vaillant2012, Pillet2016}.
In Ref. \cite{Zelener2019}, the authors chose the minimum in the steady state MOT fluorescence as the line center, estimating their uncertainty in the line center to be 10 MHz. They measured transitions to several $4sns~^1\mbox{S}_0$ levels with $n$ ranging from 40 to 120. Rather than using the quantum defect values from Ref. \cite{Gentile1990}, they fit their data to Eq.~\eqref{eqn:rydberg} to extract $E_{\rm IP}, \delta_0,$ and $\delta_1$. Their quantum defect values were consistent with Ref. \cite{Gentile1990}, although with larger error estimates. The ionization potential from Ref. \cite{Zelener2019}, $E_{\rm IP} =$ 49,305.91966(4)  cm$^{-1}$ = 1,478,154,284.9 ± 1.2 MHz assumes uncorrelated measurement errors and that all errors are statistical in nature. Their lower accuracy value is in good agreement with our results.

The greatest contribution to uncertainties in our measurements are from the frequency comb. Our lasers are tightly lock to the comb. Technical noise in the comb broadens the laser frequencies at the 1 MHz level. Reducing this noise would enable higher precision measurements of all other systematics in the experiment.

\end{document}